\documentclass[twocolumn,showpacs,preprintnumbers,amsmath,amssymb,prl]{revtex4}
\pdfoutput=1
\usepackage{graphicx}
\usepackage{float}
\usepackage{verbatim}
\usepackage{dcolumn}
\usepackage{bm}

\begin{document}


\newcommand{\beq}{\begin{equation}}
\newcommand{\eeq}{\end{equation}}
\newcommand{\beqa}{\begin{eqnarray}}
\newcommand{\eeqa}{\end{eqnarray}}
\newcommand{\lf}{\hfil \break \break}
\newcommand{\ahat}{\hat{a}}
\newcommand{\adag}{\hat{a}^{\dagger}}
\newcommand{\adagg}{\hat{a}_g^{\dagger}}
\newcommand{\bhat}{\hat{b}}
\newcommand{\bdag}{\hat{b}^{\dagger}}
\newcommand{\bdagg}{\hat{b}_g^{\dagger}}
\newcommand{\chat}{\hat{c}}
\newcommand{\cdag}{\hat{c}^{\dagger}}
\newcommand{\dhat}{\hat{d}}
\newcommand{\nhat}{\hat{n}}
\newcommand{\ndag}{\hat{n}^{\dagger}}
\newcommand{\den}{\hat{\rho}}
\newcommand{\phihat}{\hat{\phi}}
\newcommand{\Ahat}{\hat{A}}
\newcommand{\Adag}{\hat{A}^{\dagger}}
\newcommand{\Bhat}{\hat{B}}
\newcommand{\Bdag}{\hat{B}^{\dagger}}
\newcommand{\Chat}{\hat{C}}
\newcommand{\Dhat}{\hat{D}}
\newcommand{\Ehat}{\hat{E}}
\newcommand{\Lhat}{\hat{L}}
\newcommand{\Nhat}{\hat{N}}
\newcommand{\Ohat}{\hat{O}}
\newcommand{\Odag}{\hat{O}^{\dagger}}
\newcommand{\Shat}{\hat{S}}
\newcommand{\Uhat}{\hat{U}}
\newcommand{\Udag}{\hat{U}^{\dagger}}
\newcommand{\Xhat}{\hat{X}}
\newcommand{\Zhat}{\hat{Z}}
\newcommand{\Xdag}{\hat{X}^{\dagger}}
\newcommand{\Ydag}{\hat{Y}^{\dagger}}
\newcommand{\Zdag}{\hat{Z}^{\dagger}}
\newcommand{\Ham}{\hat{H}}
\newcommand{\bis}{{\prime \prime}}
\newcommand{\tris}{{\prime \prime \prime}}
\newcommand{\ket}[1]{\mbox{$|#1\rangle$}}
\newcommand{\bra}[1]{\mbox{$\langle#1|$}}
\newcommand{\ketbra}[2]{\mbox{$|#1\rangle \langle#2|$}}
\newcommand{\braket}[2]{\mbox{$\langle#1|#2\rangle$}}
\newcommand{\bracket}[3]{\mbox{$\langle#1|#2|#3\rangle$}}
\newcommand{\mat}[1]{\overline{\overline{#1}}}
\newcommand{\dotp}{\mbox{\boldmath $\cdot$}}
\newcommand{\tp}{\otimes}
\newcommand{\op}[2]{\mbox{$|#1\rangle\langle#2|$}}
\newcommand{\hak}[1]{\left[ #1 \right]}
\newcommand{\vin}[1]{\langle #1 \rangle}
\newcommand{\abs}[1]{\left| #1 \right|}
\newcommand{\tes}[1]{\left( #1 \right)}
\newcommand{\braces}[1]{\left\{ #1 \right\}}



\title{Synthesis of arbitrary, two-mode, high visibility $N$-photon interference patterns}
\author{Saroosh Shabbir}
\affiliation{Department of Applied Physics, Royal Institute of Technology (KTH)\\
AlbaNova University Center, SE - 106 91 Stockholm, Sweden}
\author{Marcin Swillo}
\affiliation{Department of Applied Physics, Royal Institute of Technology (KTH)\\
AlbaNova University Center, SE - 106 91 Stockholm, Sweden}
\author{Gunnar Bj\"{o}rk}
\email[e-mail:]{gbjork@kth.se}
\affiliation{Department of Applied Physics, Royal Institute of Technology (KTH)\\
AlbaNova University Center, SE - 106 91 Stockholm, Sweden}

\date{\today}

\begin{abstract}
 Using coherent states and linear optics, we demonstrate the synthesis of arbitrary interference patterns and establish that neither the shape nor the visibility of $N$-photon interference patterns can be used as a quantum signature in general. Specific examples include saw-curve and rectangle curve interference patterns, and phase super-resolution with period shortening of up to 60 times compared to ordinary interference. The former two with a visibility close to 100\% and the latter with a visibility in excess of 57 \%.
\end{abstract}


\maketitle
The rapid development of experimental techniques has led to the demonstration of many remarkable quantum interference effects. A decade ago, it was shown that with specific $N$-photon quantum states, one could break the Rayleigh diffraction limit and make optical interference patterns with a smallest feature size $N$-times smaller than with ordinary light \cite{Boto}. This discovery laid the foundation for quantum lithography, and as the name conveys, it was thought that this was manifestly a quantum feature. However, superpositions and interference also manifest themselves in the classical world. It is therefore of interest to delineate what interference effects belong to the realm of the classical world, and which require quantum states.

In different contexts limits arise for how large a visibility one can obtain using classical and quantum light. For instance, in the two-photon Hong-Ou-Mandel experiment, one can in principle achieve 100 \% visibility with with both the input state $\ket{1}\otimes\ket{1}$ and the state $\ket{1}\otimes\ket{\alpha}$, where $\ket{\alpha}$ is a weak coherent state. However, two mutually phase randomized classical input states will never reach a visibility in excess of 50 \% \cite{Rarity}. Likewise, letting two classical states interfere in a Mach-Zehnder interferometer and measuring the probability of detecting $m$ and $N-m$ photons respectively in the two output ports will likewise never result in a visibility of the $\lambda/N$-period fringes $> 50$ \% \cite{Afek 2}. For three and four photon visibility experiments other limits to the obtainable visibility are 81.8 \% and 94.4 \%, respectively \cite{Ficek,Agafonov}. However, all these limits are derived with the reference to a specific setup and a specific detection method. Here, we consider the interference between two coherent states using a general $N$-photon projection measurement (realized with linear optics and coincidence detection). In this case we show that there is no difference between classical and quantum states neither regarding the visibility (that can be 100 \% in both cases) nor in the obtainable shape of the interference curve. Only by restricting the measurements to $\ket{m,N-m}\bra{m,N-m}$-projectors one will regain the results in \cite{Afek 2}. The distinguishing quantum signature is instead in success probability. Only with quantum states can one surpass limits to, e.g., phase sensitivity of classical light \cite{Nagata,Okamoto}. This is regardless of the detection method.

A general, two-mode, $N$-photon state can be written
\begin{eqnarray}
\ket{\psi} & = & \sum_{n=0}^{N} c_n \ket{n,N-n} \nonumber \\
&& = \sum_{n=0}^{N} b_n (\adag)^{n} (\bdag)^{N-n}\ket{0,0},
\label{Eq: Square wavefunction} \end{eqnarray}
where $b_n = c_n/ \sqrt{n! (N-n)!}$. If we formally divide $\sum b_n (\adag)^{n} (\bdag)^{N-n}$ with $(\bdag)^N$ and make the substitution $(\adag/\bdag)^n = z^n$ we get the complex polynomial
\beq
\sum_{n=0}^{N} b_n z^{n} = b_N(z-z_1)(z-z_2) \cdots (z-z_N),
\label{Eq: Polynomial}\eeq
where we have used the fact that any complex $N$-th degree polynomial has exactly $N$ complex roots. Hence, it is always possible to express any two-mode, $N$-photon state
\beq
\ket{\psi} = \frac{b_N}{\Pi_{n=1}^N {\cal N}_n}{\cal N}_1(\adag-z_1 \bdag)\cdots {\cal N}_N(\adag-z_N \bdag)\ket{0,0},
\label{Eq: Single projector expansion}\eeq
where ${\cal N}_n$, are real numbers such that ${\cal N}_n^2(1+|z_n|^2)=1$. Thus, any two-mode, $N$-photon state can be written as a direct product of two-mode single photon states. Hofmann \cite{Hofmann} made the important observation that to make a probabilistic projection measurement onto $\ket{\psi}\bra{\psi}$, one could split the state to be measured into $N$ two-mode paths and using linear optics, unitarily transform the state in path $n$ as ${\cal N}_n(\adag-z_n \bdag)\ket{0,0} \rightarrow \adag_n \ket{0,0}$. Then, in the case one detects one photon in the $a_n$ mode of each path, one has in fact projected the input state onto $\ket{\psi}\bra{\psi}$. It is convenient to use polarization states and, e.g., take the two modes $a_n$ and $b_n$ to be horizontally and vertically polarized modes. Then any single photon polarization transformation can be achieved with two wave plates: The first compensating the relative phase between the $a_n$ and $b_n$ mode by $\theta_n=-\textrm{Arg}(z_n)$ so that the (in general elliptically polarized) state ${\cal N}_n(\adag_n-z_n \bdag_n)\ket{0,0}$ is transformed to the linearly polarized state ${\cal N}_n(\adag_n-|z_n| \bdag_n)\ket{0,0}$. The second being a polarizer rotated to the angle $\varrho_n = \arctan |z_n|$ so that the linearly polarized state passes it, but the orthogonal polarization is blocked.

This method was further developed in \cite{Guo} for measurement projection of NOON states, and demonstrated in \cite{RP,Kothe} to project out the NOON state components from coherent states to demonstrate $\lambda/N$-period interference. In \cite{Sun} the method was shown to provide a means of breaking the Heisenberg limit in interferometry when used with a $\ket{N/2,N/2}$ state input. A similar method was used in \cite{Zanthier,Oppel} to demonstrate resolution beyond the Rayleigh limit using independent light sources, such as thermal light. A scheme to synthesize arbitrary filter functions from $N$ cascaded polarization ``components,'' based on a similar factorization was suggested \cite{Harris} in 1967. However, in this proposal the individual components  were frequency modes and not single photon states. 

The stated factorization method can be taken further. Suppose the input state is a two-mode, linearly polarized coherent state
\beq\ket{\alpha,\alpha}=\exp(-|\alpha|^2)\sum_{m=0}^\infty \sum_{n=0}^\infty \frac{\alpha^{m+n}}{\sqrt{m! n!}}\ket{m,n},\eeq where for simplicity we take $\alpha$ to be real. We want to detect the interference between the phase-shifted input state and the projector $\ket{\psi}\bra{\psi}$ where $\ket{\psi}$ is given by Eq. (\ref{Eq: Square wavefunction}). The unitary, differential phase-shift operator is $\hat{U}(\phi) = \exp[-i \phi(\adag \ahat - \bdag\bhat)/2]$. The detected $N$-photon coincidence probability then becomes
\begin{eqnarray} P(\phi)& \propto & |\bra{\alpha,\alpha}\hat{U}^\dagger(\phi)\ket{\psi}|^2 \nonumber \\
& = & \alpha^{2N} \exp(-2 \alpha^2)\left |\sum_{n=0}^N b_n \exp(i \phi [2n-N]/2)\right |^2 \nonumber \\
& = & \alpha^{2N} \exp(-2 \alpha^2)\left |\sum_{n=0}^N b_n \exp(i \phi n)\right |^2.\label{Eq: Arbitrary expansion}\end{eqnarray}
The sum within the absolute sign can be identified as a truncated Fourier series. Therefore, up to the highest ``frequency'' $N \phi/2$ in the parameter $\phi$, any function can be expanded in the exponential function basis. This means that we can mimic any $N$-photon state's projection onto any $N$-photon projector with a coherent state input!

\begin{figure}[ht]
\includegraphics[scale=0.35]{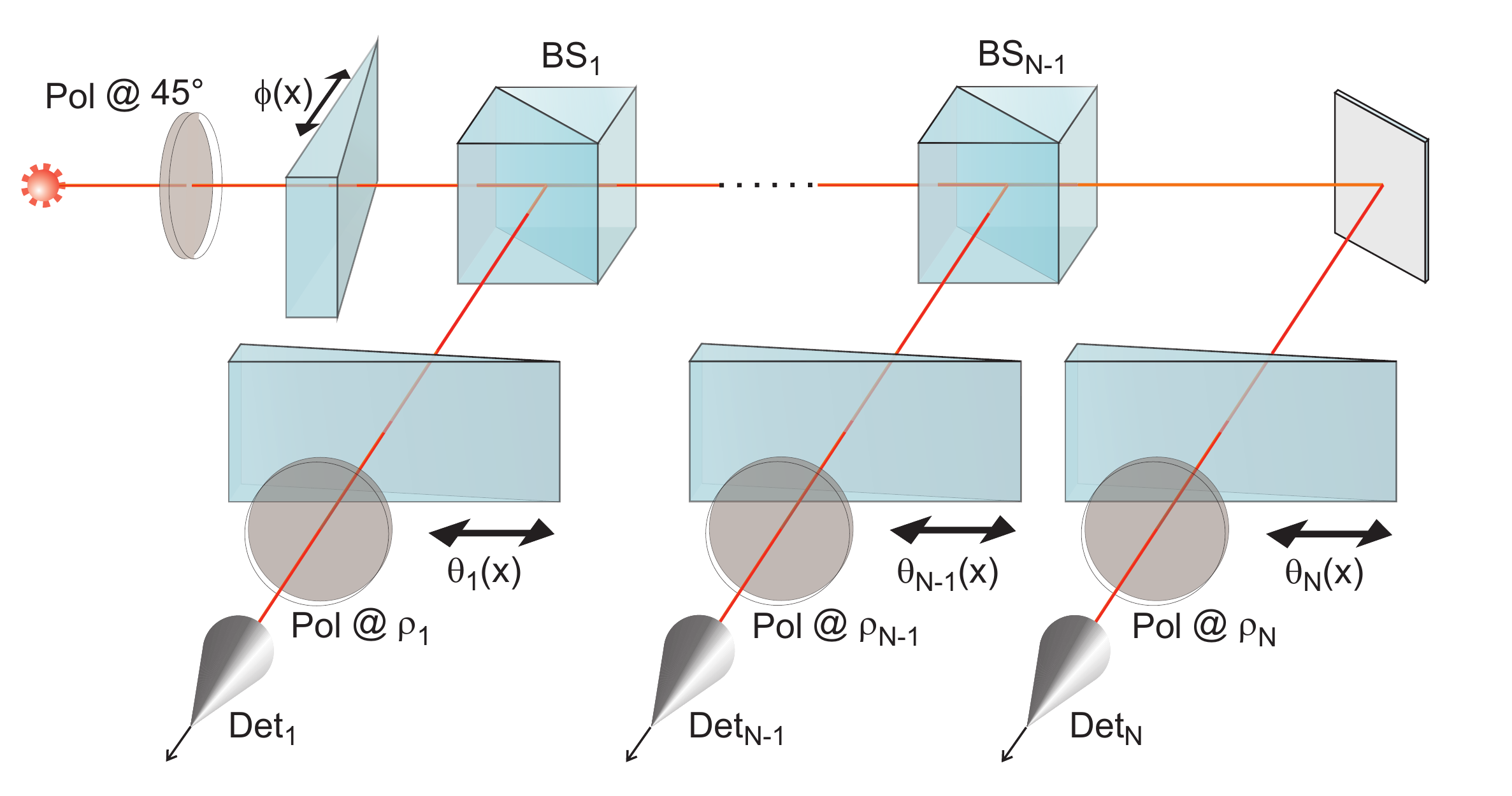}
\caption{Projection measurement setup. A weakly excited two-mode coherent state, linearly polarized at 45 degrees from the vertical, impinges from the left a birefringent wedge imparting the differential phase shift $\phi$. It is subsequently divided by a sequence of non-polarizing beam splitters into $N$ paths. In each path, a certain single photon state is projected out by preceding the detector with a polarizer set at $\varrho_n$ and a birefringent wedge set to the differential phase $\theta_n$. Finally the detection coincidences are recorded.}
\label{Fig: Setup}
\end{figure}

However, splitting a polarization, two-mode coherent state $\hat{U}(\phi)\ket{\alpha,\alpha}$ spatially into $N$ paths results in a product state of $N$ identical coherent states $\hat{U}(\phi)\ket{\alpha/\sqrt{N},\alpha/\sqrt{N}}$. The fact that this is a product state makes each state's projection probability onto a certain single photon projector, v.i.z.
\beq
P_n(\phi) = |\bra{\alpha/\sqrt{N},\alpha/\sqrt{N}}\hat{U}^\dagger(\phi){\cal N}_n(\adag-z_n \bdag)\ket{0,0}|^2
\eeq
statistically independent of any other coherent state's projection probability \cite{Glauber}. The total probability is simply the product $P(\phi) = \Pi_{n=0}^N P_n(\phi)$ of the $N$ individual projection probabilities. Thus, if one uses coherent input states it is not necessary to measure the ``clicks'' in each path in coincidence. Instead, each single photon projector $n$ can be measured separately using a weak coherent state. The result $P_n(\phi)$ is recorded, and the final probability $P(\phi)$ is subsequently obtained by multiplying all the individual probabilities as demonstrated for a NOON state in \cite{Kothe}. Hence, we need only to implement one of the arms in Fig. \ref{Fig: Setup} at a time, and reconfigure the bifrefringence $\theta_n$ and polarizer angle $\varrho_n$ between each run. Note that this is only possible for coherent states. If, e.g., quantum states or thermal states are used, as is the case in \cite{Sun} and \cite{Oppel}, respectively, there are correlations between the different single photon projection probabilities and a coincidence measurement is a must, whereas for coherent states either method works.

\begin{figure}[ht]
\includegraphics[scale=0.8]{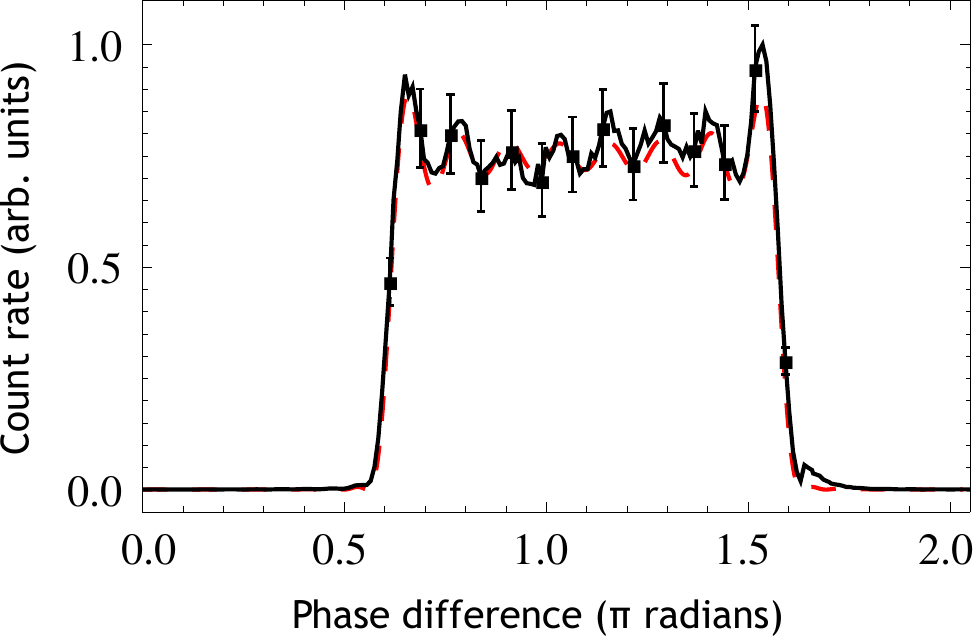}
\caption{Rectangular interference pattern. The solid (black) line (215 data points) shows raw data and the dashed (red) line shows the theoretically expected curve with the amplitude chosen for best fit. At selected data points, error bars show the $\pm \sigma$ statistical uncertainty.}
\label{Fig: Rect}
\end{figure}

To demonstrate the generality of the method, we first synthesized a rectangle function interference pattern. That is, we would like to have
\beq |\bra{\alpha,\alpha} \hat{U}(\phi)\ket{\psi_s}|^2 \propto {\rm Rect}(\phi, \pi/2) = \left \{ \begin{array}{ll} 0 & |\phi| \leq \pi/2,\\ 1 & \mbox{otherwise.} \end{array} \right . \eeq To this end, for $N=30$ which gives a reasonable rectangle-like function, we compute the 31 lowest Fourier expansion coefficients of the Rect function and identify the coefficients with the expansion coefficients $b_n$ in Eq. (\ref{Eq: Arbitrary expansion}). Subsequently the expansion is factored in single photon projector terms by Eqs. (\ref{Eq: Polynomial}) and (\ref{Eq: Single projector expansion}) (see Supplementary Material). The
thirty single photon projectors are then measured and multiplied. The result is displayed in Fig. \ref{Fig: Rect} 
where the only fitting is the amplitude of the theoretically predicted coincidence probability. The experimental curve represents raw data, no background subtraction or any other data processing have been done. The maximum count rate employed in the experiments was around 5.2 MHz. A discussion about the error bars will follow below.

\begin{figure}[ht]
\includegraphics[scale=0.8]{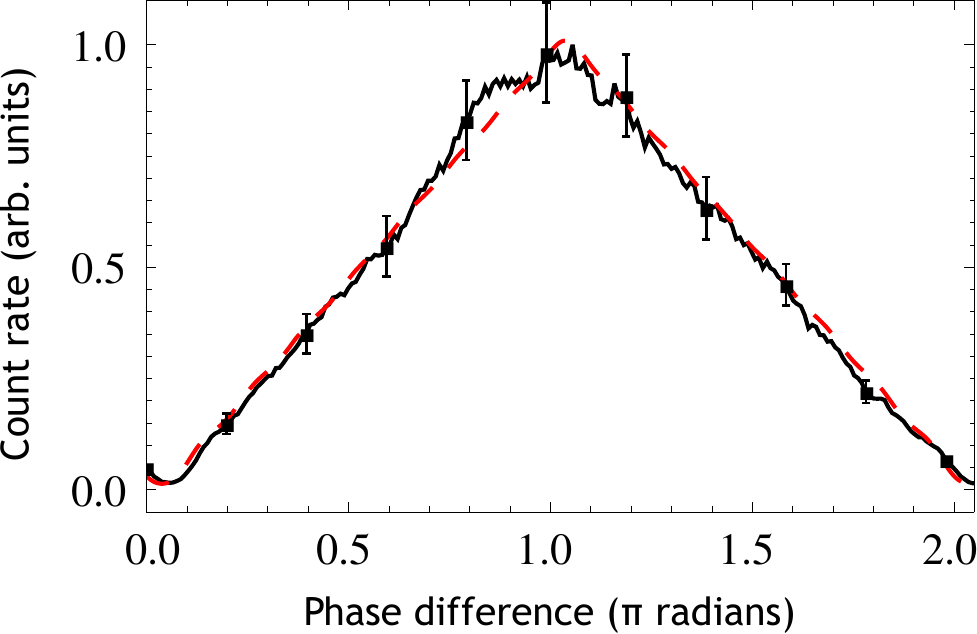}
\caption{Saw-curve interference pattern. The solid (black) line shows raw data and the dashed (red) line shows the theoretically expected curve with the amplitude chosen for best fit. At selected data points, error bars show $\pm \sigma$ statistical uncertainty.}
\label{Fig: Saw}
\end{figure}

Next, we synthesized a Saw-curve interference pattern. The interference pattern being the square of the overlap between the input state and the projector, we want the ($2 \pi$-periodic) overlap between the input state and the measurement projector to be \beq |\bra{\alpha,\alpha}\hat{U}(\phi)\ket{\psi}|=|\phi/\pi|^{1/2} \eeq in the interval $\{-\pi,\pi\}$. Again finding the Fourier coefficients for $N=30$, plugging them into Eq. (\ref{Eq: Polynomial}), and factorizing the polynomial to obtain the single photon projectors (see Supplementary Material), we finally get the raw data displayed in Fig. \ref{Fig: Saw} where the only fitting is the amplitude of the theoretically predicted curve. This interference pattern has the particular feature that its derivative, which governs the phase sensitivity, is almost constant over large intervals.

\begin{figure}[ht]
\includegraphics[scale=0.45]{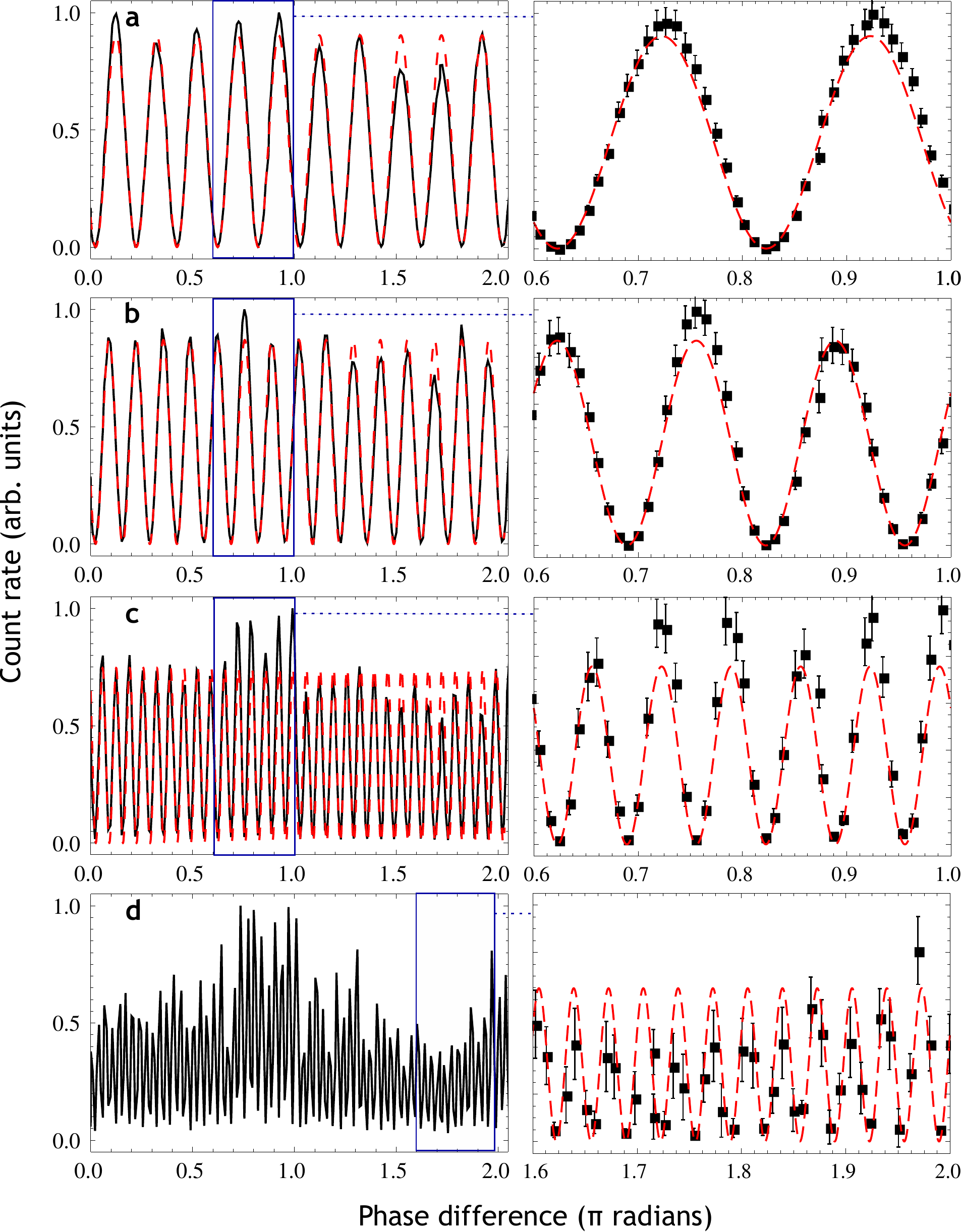}
\caption{NOON-state coincidence patterns. In the figures on the left, the solid (black) lines represent raw data points connected by straight lines. No background subtraction has been done. The dashed (red) curves are the expected $A\sin^2\left(N\phi/2 \right)$ with $A$ chosen for best fit. The figures on the right show a magnified part of the data points with error bars points and the fit curve. Error bars show $\pm \sigma$ statistical uncertainty. \textbf{a,} $N=10$. \textbf{b,} $N=15$. \textbf{c,} $N=30$. \textbf{d,} $N=60$. The boxed area shows the portion with highest visibility, with the maximum of 86.2 \% occurring at 1.75 $\pi$.}
\label{Fig: NOON}
\end{figure}

Finally, a NOON state
\beq
\ket{\psi_{\textrm{NOON}}} = \frac{1}{\sqrt{2}} \left ( \ket{N,0} - \ket{0,N}\right ),
\eeq
can give an interference pattern with the smallest feature size $\approx\lambda/(2 N)$, where $\lambda$ is the wavelength of the light \cite{Boto}. This feature size reduction can be explained in terms of maximizing a state's dynamical evolution speed \cite{Margolus,Sanders,Soderholm,Noon4} or in terms of $N$-photon quasi-particles having $N$ times the linear momentum of the photons making up the quasi particles, and thus having a de Broglie wavelength $N$ times smaller \cite{Jacobson,Fonseca,Edamatsu,WP}. Up to now, there have been many proposals and demonstrations \cite{Rarity,Fonseca,Edamatsu,WP,Angelo,Mitchell,RP, Kothe,Afek,Israel,Guo} of NOON-state interference with $N=2$ to $N=30$. The $N=6$ to $N=30$ experiments \cite{RP,Kothe}, were made with coherent light with the same method we now generalize. There has also been other demonstrations and proposals of multi-photon (i.e., using coincidence measurements) interference using non-entangled, classical states also showing a period shortening, but at the expense of a reduced interference visibility. In, e.g., \cite{Afek 2} periods down to 1/4 of the ``regular'' interference period were measured, but with a visibility of only about 29 \%. Using the method above, we obtained interference patterns for projection of a coherent state onto $N=10$, $15$, $30$, and $60$ NOON state projectors, shown in Fig. \ref{Fig: NOON}. We see that for $N=10$ we get an excellent fit with the predicted pattern. The poorest visibility, where the visibility of a fringe is defined by the adjacent local minimum and local maximum of the interference pattern is 99.5 \% in this case. Even for $N=60$ we get a reasonable fit with the theory and a minimum and maximum visibility of 57.5 \% and 86.2 \%.

The visibility is limited by five effects. The first one is the smallest step with which we can vary $\phi$. It is set by the combination of birefringent material (quartz in our case), the birefringent wedge angle which is 19 deg, and the linear motorized stage minimum step size 1 $\mu$m. This leads to a minimum phase difference resolution of 29 mrad which for $N=60$ is too poor (one period of the pattern, occupying $2 \pi/60$ radians, is probed by less than four measurement points). As seen in Fig \ref{Fig: NOON} (d), we simply ``miss'' many minima and maxima of the pattern, yielding a lower visibility than that set by other measurement errors and noise. For $N \leq 30$, the dominant source of error is the stochastically varying quantum efficiency of the detector. The measured coincidence probability is $P(\phi)= \Pi_{n=1}^N\eta_n P_n(\phi)$, where $\eta_n$ is the quantum detector efficiency for the single photon projector $n$. In our case, we use the same detector for each projector, and therefore $\eta_n \rightarrow \eta(n \tau)$, where $\tau = 480$ s is the time it took between measuring the point $\phi$ of one projector to the same point for the next projector. Assuming that the variation in quantum efficiency $\Delta \eta(n \tau) = \eta(n \tau)-\langle \eta(\tau)\rangle$ is uncorrelated between one projector to another, the expected standard deviation due to this error is
\beq \langle (\Delta P(\phi))^2 \rangle^{1/2} = \sqrt{N} P(\phi) \frac{\langle \left [ \Delta \eta(n \tau)\right ]^2 \rangle^{1/2}}{\eta} .\eeq
This equation shows two important features. Firstly, that due to the multiplication of detection events the error will grow with the square root of the projector's photon number. This sets an upper limit to the method's applicability. Secondly, the absolute error is proportional to the coincidence probability, or stated otherwise, the relative error is constant. This is clearly seen in our data as an excellent fit between experiment and theory as long as $P(\phi)$ is small. Another error source, but less important for our measurements, is small errors in the phases $\theta_n$. For a rapidly oscillating function, the exact position of each single photon projector's minimum has a substantial impact, as predicted in \cite{Kok}. In addition, for the NOON state interference, one should ideally be able to get all the light in and out of one linear polarizer. At best, we manage a perpendicular polarization suppression ratio of 46 dB, and typically the suppression rate is $> 40$ dB. Finally, the detector dark count also sets a limit. For our NOON state interference this error source is negligible as the dark count rate is $< 2$ Hz and the maximum count rate is 1.02 MHz. The detector efficiency (if it were constant) does not matter in the experiment as the detected state is a coherent state irrespective of $n$ and $\phi$. A low detector efficiency (about 20 \% for our detector) can easily be offset by a higher input intensity.

With the Hofmann-Guo method \cite{Hofmann,Guo}, it is possible to synthesize any interference pattern to, in principle, the precision given by $N$. The method does not limit the obtainable visibility, and we have shown that in practice, even $N=60$ NOON-state interference patterns can be recorded with $>50$ \% visibility. We have also demonstrated that unusual (non-sinusoidal) interference patterns can arise from multi-photon interference of coherent states. Hence, it is neither the shape nor the visibility of multi-photon interference patterns that delineate a general border between classical and quantum interference, although in special cases limits do apply.

However, with a coherent state input it is not possible to get quantum phase sensitivity \cite{Mitchell,Nagata,Okamoto} (i.e. break the standard quantum limit \cite{Caves,GL}). On the contrary, as discussed in some detail in \cite{Kothe}, for a given mean photon number, e.g., the NOON state interference curves have a lower phase sensitivity than ordinary (single fringe) interferometry with a coherent state.

It is possible to directly scale up the demonstrated method to the classical regime using ordinary photo detectors, and perhaps it is here the method will find applications. One reason we have stuck to working at the single photon detection level is to show that the coincidence method is very flexible in that allows the synthesis of \textit{any} two-mode, multi-photon projector, albeit at the expense of exponentially decreasing probability of coincidence events with increasing $N$. Another reason is to explore what practical limits (i.e., how the influence of various sources of errors) one encounters when one scales single-photon coincidence measurements to large numbers.

When using coherent state input, the switch from obtaining the interference curves by coincidence measurements to multiplication of single photon projection probabilities allows considerable savings in time and equipment. In our case, the maximum probability of detecting a photon in a single temporal mode, defined by the response time of the detector, is about 0.1 for each single photon projector. Hence, as the $N$ single photon projectors give statistically independent results, the probability of detecting, say, 10 photons in coincidence would be around $0.1^{10} = 10^{-10}$. With our detectors with a response time of 45 ns it would thus take at least $45\cdot 10^{-9} \cdot 10^{10}=450$ s to get a single coincidence click, on average. Detecting each (temporal) mode sequentially, instead of detecting $N$ (spatial) modes in parallel, it took about 900~s to measure the 60 projectors for a single phase shift $\phi$. Thus, taking 215 steps to cover the range of phase shifts from 0 to $2 \pi$ radians, a whole $N=60$ interference curve was obtained in about 8 h which is long, but feasible.

The authors thank C. Kothe, S. Takeuchi, K.~Edamatsu, and S. Harris for helpful discussions. The work was supported by the Swedish Research Council (VR) through grant 621-2011-4575 and through its support of the Linn\ae us Excellence Center ADOPT.

\newpage

\section*{Supplementary material}\label{SM}

\subsection{Methods}

We have implemented a multi-photon experiment using a linearly polarized HeNe laser (whose polarization was further ``cleaned'' with a polarizer) with a power of 15 mW, neutral density filters, two birefringent wedges mounted on step-motor drive translators (with a precision of 1 $\mu$m) a polarizer mounted on a rotation stage (with a precision of 0.2 degrees), and a single photon  sensitivity avalanche photo diode (APD), see Fig. \ref{Fig: Setup}. The intensity of the laser was adjusted by the neutral filters such that the mean time interval between two photons was 20 times longer ($\approx 1 \mu$s) than the dead time of the detector (45 ns). Thus, we ensure that essentially each recorded ``click'' stems from a single photon. The laser's polarization was carefully adjusted to be at 45 degrees from the horizontal, thus creating the desired input state $\ket{\alpha,\alpha}$, where $\alpha \ll 1$ in the temporal mode basis of the detector. The calibration of the wedges was found to give a $2 \pi$ phase shift with a relative translation of 0.215 mm.

Following the method in \cite{Kothe}, each arm of the interferometer was implemented sequentially and the results were recorded on the computer. For each arm $n$, the first birefringent wedge was set to impart the differential phase shift $\phi$ (from $0$ to $2 \pi$ for each projector) on the input state, the second birefringent wedge imparted the differential phase shift $\theta_n$ and the polarizer after the second wedge was rotated to the angle $\varrho_n$ to adjust the relative amplitude of the horizontal and vertical modes (see Fourier Expansion Coefficients). The state was subsequently focused on the APD and the number of ``clicks'' during one second was recorded as a function of $\phi$ in a computer memory. After all $N$ single photon projector counts [proportional to $P_n(\phi)$)] had been measured as a function of $\phi$, the counts at each setting of $\phi$ were multiplied together yielding a function proportional to $P(\phi)$.

All the displayed data is raw, that is, without the subtraction of dark counts and it was obtained with a counter gate time of 1 s.

\subsection{Fourier expansion coefficients}
 Consider a $2 \pi$-periodic rectangle function defined by the two conditions \beq {\rm Rect}(\phi, \pi/2) = \left \{ \begin{array}{ll} 0 & |\phi| \leq \pi/2,\\ 1 & \mbox{otherwise.} \end{array} \right ., \eeq and ${\rm Rect}(\phi, \pi/2)={\rm Rect}(\phi + 2 l \pi, \pi/2)$ where $l$ is an arbitrary integer. Since the Rect function is even, it is advantageous (and natural) to use an even number of projectors to synthesize the function implying that $N$ should be chosen even. (The function can also be expanded for odd $N$ but with a less pleasing result). The Fourier expansion of this function can hence formally be written
\beq {\rm Rect}(\phi, \pi/2) = \sum_{n=-N/2}^{N/2} b_{(2n+N)/2} \exp(i \phi n )\eeq  where the expansion coefficients are
\begin{eqnarray} b_{(2n+N)/2} & = & \frac{1}{2 \pi}\int_{- \pi}^{\pi} {\rm Rect}(\phi, \pi/2) \exp(-i \phi n ) d\phi \nonumber \\
& = &
\left \{ \begin{array}{ll} 1/2 & n = 0, \\
-\frac{\sin(n \pi/2)}{n \pi} & n \neq 0,  \end{array} \right .\\
\end{eqnarray}
Using Eq. (\ref{Eq: Square wavefunction}), the associated polynomial for this state for, e.g., $N=10$ is hence \beq \frac{z^5}{2}-\frac{z^4 + z^6}{\pi}+\frac{z^2+z^8}{3 \pi} -\frac{1+z^{10}}{5\pi} = 0, \eeq with the roots
\begin{eqnarray}
 z_1 = z_2^* & = & 0.967612 + i 0.252442, \nonumber \\
 z_3 = z_4^* & = & 0.723141 + i 0.6907, \nonumber \\
 z_5 = z_6^* & = & 0.313207 + i 0.949685, \nonumber \\
 z_7 = z_8^* & = & -0.463687 + i 0.29332, \nonumber \\
 z_9 = z_{10}^* & = & -1.54027 + i 0.974347. \end{eqnarray}
 The fact that the Rect function is an even function leads to the result that all the roots of the associated polynomial come in complex conjugate pairs. Inserting these roots into Eq. (\ref{Eq: Single projector expansion}) it is evident how to implement the ten single projectors. For example, the $n=1$ projector is implemented by introducing a birefringence of $-\textrm{Arg}(z_1)\approx -0.255$ rad $= -14.6$ degrees between the horizontal and vertical directions. This birefringence should be followed by a polarizer set at the angle $\arctan|z_1|=\pi/4$ rad. In Table \ref{Table 1} we list the settings of the birefringence and the polarizer angle for the 10 projectors in degrees.
 \begin{table}[h]
\begin{tabular}{|c|c|c|}
  \hline
  Root $n$& $\varrho_n$ & $\theta_n$ \\ \hline
  1 & 45.0 & -14.6 \\
  2 & 45.0 & 14.6 \\
  3 & 45.0 & -43.7 \\
  4 & 45.0 & 43.7 \\
  5 & 45.0 & -71.7 \\
  6 & 45.0 & 71.7 \\
  7 & 28.7 & -147.7 \\
  8 & 28.7 & 147.7 \\
  9 & 61.2 & -147.7 \\
  10 & 61.2 & 147.7 \\
  \hline
\end{tabular}
\caption{The experimental parameters for a $N=10$ Rect-function projectors.}
\label{Table 1}
\end{table}
It is  seen in Fig. \ref{Fig: Rect} that the resulting function is not a perfect rectangle, but Gibbs phenomenon makes the interference curve overshoot on the steep flanks. This effect can be reduced by the use of a Lanczos-Fourier expansion or a Ces\`{a}ro approximation of the Fourier series, at the expense of getting less steep flanks in both cases. Going to higher photon numbers one could in principle reduce the wiggles in the interval $\{ \pi/2, 3 \pi/2 \}$, but the overshoot height would remain the same, only the width of the overshooting peak could be decreased.

\begin{table}[t]
\begin{tabular}{|c|c|c|}
  \hline
  Root $n$& $\varrho_n$ & $\theta_n$ \\ \hline
  1 & 56.2 & 0.0 \\
  2 & 33.8 & 0.0 \\
  3 & 66.6 & -74.0 \\
  4 & 66.6 & 74.0 \\
  5 & 23.4 & -74.0 \\
  6 & 23.4 & 74.0 \\
  7 & 21.7 & -141.7 \\
  8 & 21.7 & 141.7 \\
  9 & 68.3 & -141.7 \\
  10 & 68.3 & 141.7 \\
  \hline
\end{tabular}
\caption{The experimental parameters for a $N=10$ Saw-function projectors.}
\label{Table 2}
\end{table}
In a similar manner, an $N=10$ expansion of the Saw$^{1/2}$ function
\beq {\rm Saw}^{1/2}(\phi, 2 \pi) = \sum_{n=-N/2}^{N/2} b_{(2n+N)/2} \exp[-i \phi n ].\eeq Computing the expansion coefficients in the same manner as before, one arrives at  \beq b_{(2n+N)/2} = \frac{
 \sqrt{2 n}\sin(n \pi) -{\rm FS}(\sqrt{2 n})}{\sqrt{2} \pi n^{3/2}},\eeq where FS is the Fresnel sine function. Numerically, the coefficients are: \begin{eqnarray} b_5 & = & 2/3, \nonumber \\ b_4=b_6 & = & -0.1607, \nonumber \\ b_3=b_7 & = & -0.0273, \nonumber \\ b_2=b_8 & = & -0.0272, \nonumber \\ b_1=b_9 & = & -0.0109, \nonumber \\ b_0=b_{10} & = & -0.0121. \end{eqnarray}

The polynomial associated to the expansion is thus \begin{eqnarray} \frac{2 z^5}{3}- 0.1607 (z^4 + z^6)-0.0273(z^3 + z^7)  && \nonumber \\- 0.0272(z^2 + z^8)-0.011 (z + z^9) && \nonumber \\ - 0.012(1 + z^{10}) & = & 0 \hfill \end{eqnarray} with the roots
\begin{eqnarray}
z_1 & = & 1.49215, \nonumber \\
z_2 & = & 0.670175, \nonumber \\
z_3 = y_4^* & = & 0.637336 + i 2.22359, \nonumber \\
z_5 = y_6^* & = & 0.119116 + i 0.415582, \nonumber \\
z_7 = y_8^* & = & -0.311654  + i 0.245907, \nonumber \\
z_9 = y_{10}^* & = & -1.97752+ i 1.56034. \end{eqnarray}
The settings, in degrees for the birefringence and the polarizer angle are given in Table \ref{Table 2}.

\end{document}